\documentclass[12pt]{article}
\usepackage{graphicx}
\usepackage{amssymb,amsmath,amsfonts,palatino,amsthm}
\usepackage{amssymb}
\usepackage{epstopdf}
\usepackage{slashed} 
\newcommand{\f}{\begin{equation}}
\newcommand{\ff}{\end{equation}}

\begin{document}
\title{The path integral in energetic causal set models of the universe}
\author{Lee Smolin\thanks{lsmolin@perimeterinstitute.ca} 
\\
Perimeter Institute for Theoretical Physics,\\
31 Caroline Street North, Waterloo, Ontario N2J 2Y5, Canada}
\date{\today}
\maketitle
\begin{abstract}

I study several aspects of the path(st) integral we formulated in previous papers on energetic causal sets with Cortes and others.
The focus here is on quantum field theories, including the standard model of particle physics.

I show that the the theory can be extended to a quantum field theory,  cut off in momentum space.    Fields of spin 
$0, \frac{1}{2}. $ and $1$ may be included as in a perturbative treatment of the standard model.      

The theory is at first formulated in momentum space.      Under  certain conditions, spacetime can emerge in a semiclassical limit.  
The theory comes with a $uv$ cutoff in momentum space, $\mu$, hence that is also a scale for  lorentz invariance to break down.    
Traditionally,  $\mu$ is taken to be. a Planck energy, but we 
explore as a possibility making $\mu $ smaller.


\end{abstract}
\tableofcontents



\section{Introduction} We continue the series of papers on the energetic causal set approach to quantum foundations , quantum gravity and
cosmology\cite{ECS1}-\cite{VVQM}.   Our focus in this paper is to extend the range of the theory to cut off formulations of (nearly) relativistic
quantum field theories. 

We will make several  assumptions, which shape our treatment of the theory.  Some of the motivation for choices made come from quantum gravity,
although the theory I present here is not (yet) a quantum theory of gravity.

\begin{itemize}

\item{}Usually we define a quantum gravity theory by fixing a uv cutoff, $\mu$ and then looking for fixed point in the space of couplings which the theory
may approach in a limit as $\mu \rightarrow \infty$.  This is thought 
to be necessary to restore symmetries or gauge invariances that are lost for fixed $\mu$.    

But, experimentally, to my knowledge, Lorentz invariance is only checked explicitly  up to
$\gamma = \frac{1}{1- \frac{v^2}{c^2}}  \approx 10^{12}\cite{howhigh,howhigh2}$.    Moreover, there is a preferred simultaneity for the universe as a whole.
These and other ideas suggest that we might formulate quantum gravity in a way in which both global and local lorentez are broken at some
intermediate energy scale,  allowing just
enough to preserve current experiments.   The fact is that there is a lot of room between the Planck scale and the weak scale, and the lower we can put $\mu$,
the more opportunity we can open up for cutoff scale quantum gravity.   The question then is then, just {\it  how low  }  can $\mu$ go?  

\item{}As proposed originally by Cohl\cite{Cohl1},  the histories of events we sum over in the model are those of the  standard model.  
There are not two sets of events, the causal set events and the standard model events. In the usual way of thinking about it, the first  set of events define the history of the quantum spacetime. Then the standard model particles and fields propagate in the background of events defied by the first set. 
But there is no need for two sets of histories, and two path integrals.
  The second set of histories, where we include the matter degrees of freedom, is already summing over a sufficient number of paths to define the spacetime.  We don't have to do that twice.




\item{} Fundamentally, our model is just defined on pieces of momentum space.  There is no spacetime, quantum or classical.  Can we presume that
spacetime instead, emerges at around the cutoff scale?  How low can we make that scale?

Classical spacetime emerges, carrying a scale which is $l_{o} = \epsilon^{-1}$.

\item{}. Our theory comes with a tiny number,   
\f
\epsilon _0 \approx    {l_{Planck}}{\epsilon}
\ff

There is no avoiding this number; after all it is in the data.  Its origin is unknown!  The hierarchy problem is re-expressed but remains unresolved.  
But let us use it in the theory somewhere it will do us some good.

If we can make the fundamental cutoff scale, somewhere between the Planck scale and  the weak  scale, we have to explain how quantum gravity gets to be so weak.    

\item{}The very tricky question is whether there are possible observations that might be made in a theory with a weak scale cutoff that could indicate
that there are not degrees of freedom far above that scale, in energy units.   This is the converse of the question of whether current experiments, made below the weak scale,
could establish the existence of modes of a smooth spacetime, far above the weak scale.   In other words could we already rule out the scenario proposed here.
Some relevant papers  are \cite{howfar,howfar2}.

\item{} The way we treat energy, momentum and spacetime in this matter is inspired by how they are treated in {\it relative locality }models
of quantum gravity phenomenology\cite{RL1,RL2}.  .   

Usually we take spacetime ${\cal M}$ as fundamental, while momentum space is an aspect of the cotangent bundle over spacetime.  
This means that there is a momentum space for every pointy of spacetime.   Here we reverse this, we take momentum space
as fundamental.   Spacetime is not defined at first, then it emerges from the tangent space of momentum space.  This means that
if several particles interact, each carrying a different momenta, they also move in different spacetimes.  
  This means that there is
a single momentum space, but there is, strictly speaking a spacetime for every point in momentum space.   In [
\cite{RL1,RL2} we showed how this works,
giving rise to a beautiful picture of quantum gravity phenomenology, emerging with the emergence of space. 

\end{itemize}


Down at the Planck scale, the gravitational terms are order unity, which is to say strongly 
coupled, whereas due to asymptotic freedom, the gauge matter couplings are practically negligible. 
So it make sense to think about a strongly coupled gravitational plasma around the plank scale and ignore matter there.  

But if the fundamental cutoff scale is closer to the weak scale than the Planck scale, then there is nothing that forces us to solve this very hard problem, of quantum gravity in the Planck scale regime.





\section{The dynamics}

\subsection{Energetic causal dynamics}

Let us first review the dynamics of the energetic causal theory, in a generalized form.

Our model is defined by a path integral, from a fixed initial state  to a fixed final state.    Those states are labeled by representations of the Poincare group,
$p_a^{cutoff} $
up to a fixed cutoff $\mu$.    Dynamics is imposed by constraints; which are represented in the path integral by
delta functionals.  

The theory thus computes probability amplitudes 
 in momentum space.
 
We preserve rotational invariance in a specified frame, which is to say; that we include in our sums over momentum modes for momentum $| p_i | \leq \epsilon$.
\f 
Z= < p_a^{i}  |  \sum_\Gamma  \Pi_{ \Gamma}   ( \int dp_c^{p^{max}}  \delta ({\cal P} (p))  \delta ({\cal C} ) 
 |   q_b^{j } >   
 \label{past}
\ff
where the constraints are
\f
{\cal P}_a^I = \sum_{k \in node K}  p_a^{k} =0
\ff
\f
{\cal C}_K^L = p_{a K}^L h^{ab} p_{b K}^L =0
\label{C=0}
\ff

It will be important to understand how the edges and vertices are produced in the above action.    

The interactions come only from the ${\cal P}_a = 0$ constraints.   To see how this happens let's remove the terms in these constraints
\begin{eqnarray}
Z^{-{\cal P} }  &= &  < p_a^{i}  |  \sum_\Gamma  \Pi_{ \Gamma}   \int dp_c^{p^{max}}   \delta ({ \cal C }) | p_a^{i}  >  
\nonumber \\
 &  = &   < p_a^{i}  |   \sum_\Gamma  \Pi_{ \Gamma}   \int dp_c^{p^{max}}   \int d{\cal N}^J_K.  e^{\imath  \sum_{J<K}  {\cal N}^K_L {\cal C}_K^L    }  | q_b^{j } >   
\end{eqnarray}

Since the ${\cal C}_J^K$ constraints act each on one momentum, there is no coupling of the different momentum to each other.
The only place interactions are introduced
in our model is  by the conservation laws imposed by delta functions in the measure of the path integrals. 

\section{Generation of  the causal process by the path integral}

Our diagrams represent a causal process.   They are embedded neither in spacetime nor in momentum space.   Each diagram contains a
partial ordering amongst its vertices,.   

\begin{center}
\begin{figure}[!h]
\includegraphics[scale=.09]{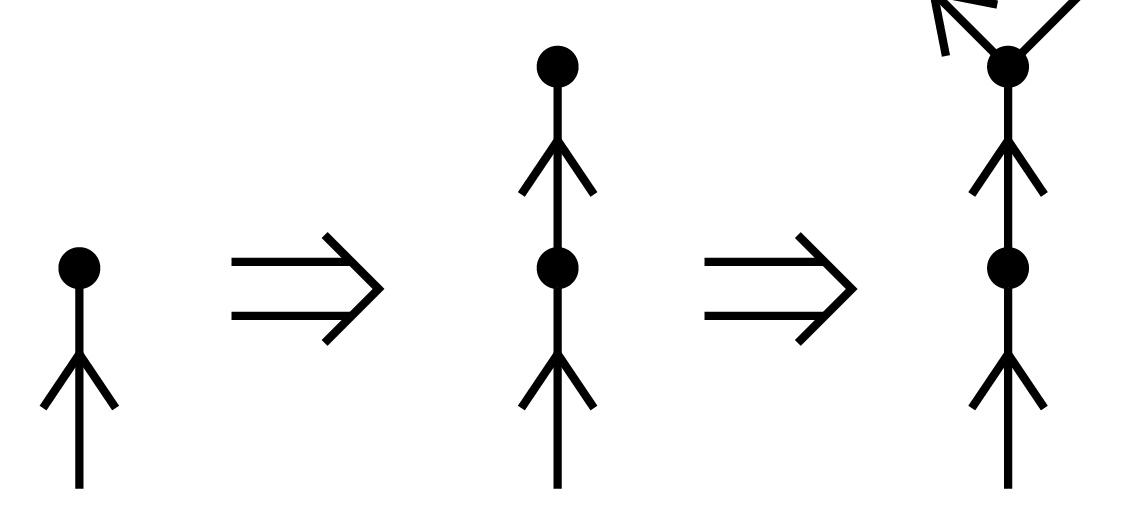}
  \caption{Two basic moves, one after another}
\end{figure}
\end{center}

\begin{figure}[!h]
\includegraphics[scale=0.09]{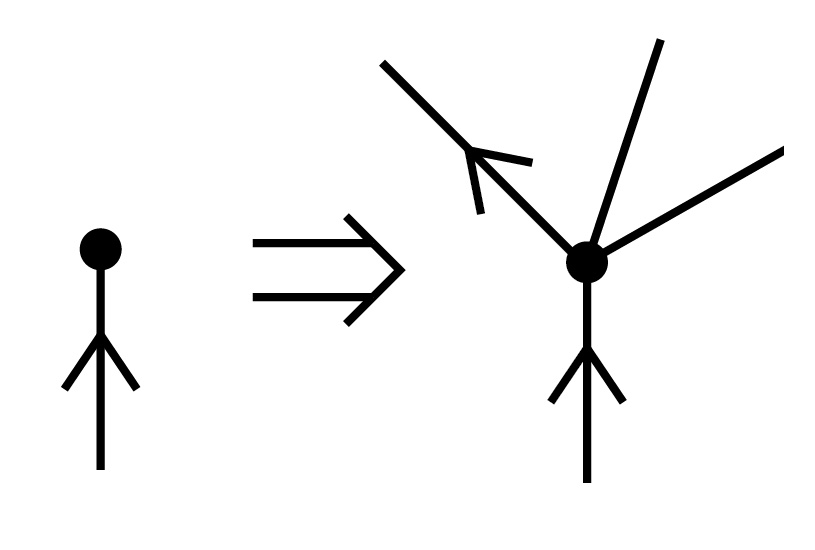}
  \caption{Another basic move.  In these diagrams we see described the partially ordered casual structure of the nodes.}
\end{figure}

\begin{figure}[!h]
\includegraphics[scale=.2]{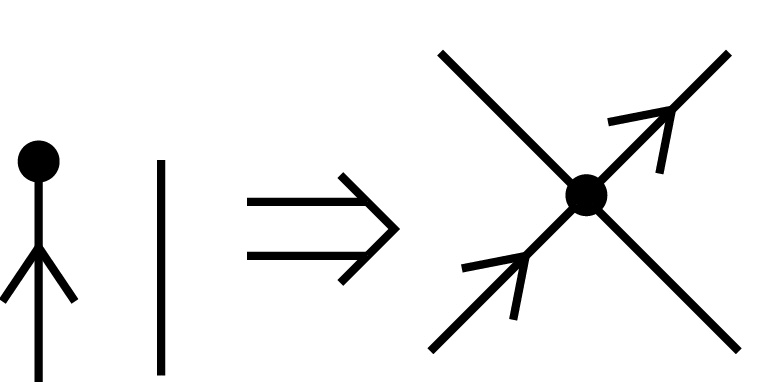}
  \caption{Another basic move.}
\end{figure}

A diagram in our theory is a contribution to the past integral \ref{past}.   We see that the diagrams are made up of nodes, each one of which has 2, 3 or 4 
inputs or outputs.  They are connected by ppropagators.  There are only two kinds of particles, which are chiral fermions and real scalar particles
In a later section, we show how to extend the  notation to incorporate all the fields of the standard model
ie lptons, quarks, W bosons, photons, gluons and Higgs
particles.        

A diagram starts off with a specific number of free particle states, which are eigenstates of the momentum operators.    Each free particle represents
a propagator, constructed by summing up an infinite number of two point functions, as in equation (\ref{sumup}) . 


\subsection{The two point function}

At each stage in the causal order, you see a static diagram, which contains, first, the initial incoming  states, then a diagram built
on them. 

Each incoming particle is represented by an appropriate line, ending at a node (these initial nodes have just a single past input each.)

After one has performed the sum
\f
{\cal J}   = \delta^I_{I+1}    +   (p_a )^2   p^2 + p_{a}^2 p^4 + \ldots = \frac{1}{1 - p^2 }
\label{sumup}
\ff

To extend the path integral () to the standard model, we need to  first derive expressions for particles to propagate.  These come  from summing up all the two
point functions. This is illustrated in Equation (\ref{sumup}). 


\subsection{Causality and the $\imath \epsilon $ rule}

There is only one place that position spacetime is referred to in the path integral, which is that we must recover the  $+ \imath \epsilon $ prescription in the
propagators, as that refers to positive frequency goes to the future and negative frequency goes to the past.  
These $+ \imath \epsilon $ factors are imposed on the Fourier transforms of our momentum space amplitudes; if we fail to replicate
the effects of this, the Feynman path integral is not reproduced.

However, the causal structure is already there in our pure momentum space factors.  The reason is that we are required to go through the path integral
in a causal order.  Our problem is just to make sure that the factors coming from this, correct ordering,  is preserved when we write out the Fourier expansion
of the expansion of our path integral (\ref{past}).
So we see that the causal structure of each amplitude is preserved in the embedding of the Fourier transformed map into spacetime.
\f
Z^{SM}    =  < p_a^{i}  |  \sum_\Gamma  \Pi_{ \Gamma}   \int dp_c^{p^{max}}  \prod_{vertices}   \delta ({ \cal P }_{a \ vert})   e^{\imath {\cal W}^{top}    } | p_a^{i}  >  
\ff


\begin{itemize}

\item{} Our next step is to extend the constraint that gives the scalar particle its transformation properties, in order to introduce the action of the $SU(2)_{Left}$ group.
Instead of a single scalar field $\Phi $ we have a doublet from the electroweak  group $H$.  This is represented by
\f
    {\cal C}^{H} = H^{\dagger}_{A'} ( p_a   g^{ab} p_b )  \eta^{A'A}  H_{A} +V
\ff
where
\f
V= -\frac{\mu^2 }{2} H_{A}  \eta^{A'A}  H_A  +    \frac{\lambda}{24}   [ H^{\dagger}_{A'}   \eta^{A'A}  H_A ]^2
\ff
is the scalar potential.

Again we sum over all two point functions for the Higgs.
A Higgs has an arrow next to its edge.
 
\item{} Next in complexity is the pure fermion, sector,  which is to start with the form,
\f
S^{\psi} = \sum_{I}^{\ K}  \Psi^{\dagger A^\prime } (p)  \sigma_{A^\prime }^{a B}  p_a  \Psi (p)_{B}
\ff
We can sum over an infinite number of two point diagrams to construct the propagator, as we see in  eq(\ref{sumup}). 
A left handed chiral fermion has an arrow on to its edge.

\subsection{The vertices}

To construct the vertices of the standard model, one needs to expand the path integral.

\begin{figure}[!h]
\includegraphics[scale=.05]{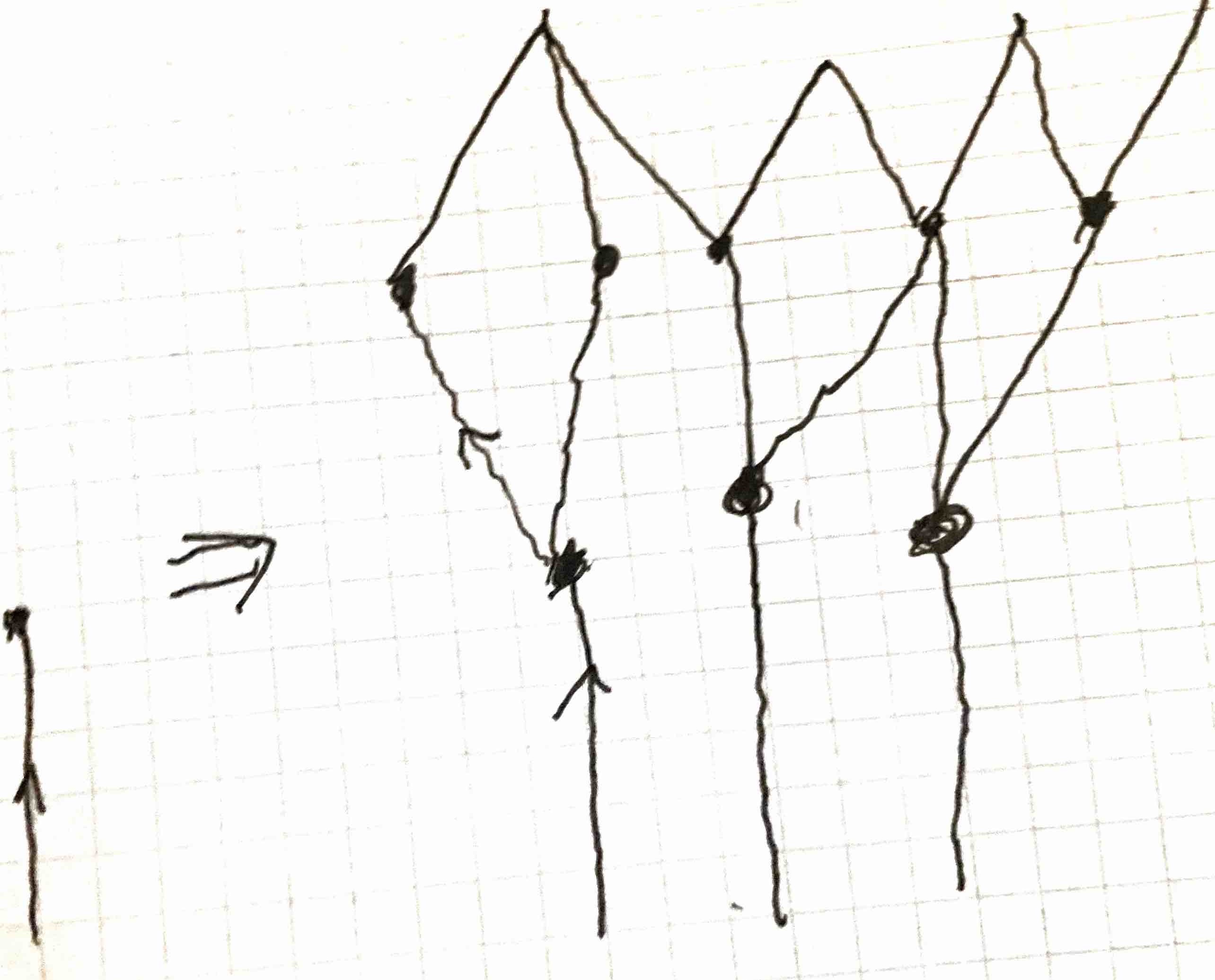}
  \caption{The first few moves in an evolution beginning with one initial chiral fermion and two bosons.}
  \label{FIg4}
\end{figure}


It is interesting also to note that locality in the emergent spacetime is a consequence of the linearity of the conservation laws. 
It is important that all these processes satisfy the energy- momentum  conservation rules; this is accomplished by inserting the 
appropriate delta functions downstairs, in the measure of the path integral.

The graph then grows in a series of steps.  In each step one of several things happen. 
\begin{itemize}

\item{} One, two or three edges are added moving to the future of a free present node.   (See Figures 1 and 2 ). 

The nodes correspond to events.  They are time ordered, in the sense that the causal, partial order is constructed according to the
partial order of adding nodes to the graphs.

\item{} Two present nodes come together and become a single node with two or more input edges.   (See Figure (\ref{Fig3} ).

\item{} Nodes are in the present if they may be built on by the assignment of further nodes to them.  


\item{}As in the expansion of the "past" integral in the original energetic causal set models, a slight modification of our
model allows us to  interpret it in the context of
a presentist view of time.  We may put a restriction on nodes to limit future growth from it, for example no more than $m$ nodes directly
to its future.  When a mode saturates this restriction, it can no longer be a pathway to the future; and we move it to a set of  {\bf past nodes}.
Once a mode is in the past, it cannot move back into the future\cite{ECS1}, \cite{ECS2}.

We feel justified in saying that only present nodes are part of the real, but we recognize this may be a matter of taste\cite{ECSinthefuture}.

In ordinary causal set models, we cannot impose such a restriction because it makes demonstrations of lorentz invariance impossible\cite{CS1}.
We have no need to preserve lorentz invariance at arbitrary distances from a given point, so we easily live with the dropping of Lorentz
invariance.

In the diagrams  drawn here there are two kinds of edges, those corresponding to chiral fermions, and those corresponding to real
scalar fields.  The chiral fermion modes never end, and they preserve a chiral fermion number.   The real scalar modes can
start or stop at a node.

\end{itemize}



In the simple model we are notating, the chiral fermion may be in one of four momentum eigenstates, which correspond to parity and charge.
while the scalar may be in one of two states.  This is sufficient to show the elementary logic behind the $CPT$ theorem.  
The theorem still works when we increase the complexity to incorporate the standard model. 

\subsection{Discrete symmetries and $CPT$ theorem.}

There are four charged states that a particle may find itself in, with respect to the present:  
$Q=1,-1, \pm= + $ or $-$.  We can write there four states like
$\frac{\pm 1}{\pm}= \frac{Q}{\pm}$.  On the other hand, if we have a charge neutral state, it only can change the direction of its arrow, so it has two states
$\pm = \frac{0}{\pm}$.   
We define the following three discrete symmetries:

Parity. ${\cal P}$:
\f
{\cal P}  Q = Q,     {\cal P}\pm = -\pm. ,  \ \ \ \ \ \ \ \ ie \  \   \ \   {\cal P}:  \frac{Q}{\pm} =\frac{Q}{-\pm}. 
\ff
Charge conjugation:  {\cal C}:
\f
{\cal C}: {\cal Q} = - {\cal Q}; \ \ \ \ \ \ {\cal C}: \pm = \pm.  \ \ \ \ \ \ \ \   ie \  \   \ \   {\cal C}:  \frac{Q}{\pm} =\frac{-Q}{\pm}. 
\ff
Time reversal {\cal T}  
\f
{\cal T}: Q = -Q , \ \ \ \ \ {\cal  T}:  \pm = \mp.   \ \ \ \   ie \  \   \ \   {\cal T} \  \frac{Q}{\pm} =\frac{-Q}{-\pm}.
\ff

It is easy to check that these transformations are not trivial on both kinds of states we are working with.

On the other hand $CTP$ is the identity on our states,
\f
{\cal T}{\cal C} {\cal P}= {\cal I}
\ff


\subsection{The gauge fields}

\item{} We next introduce the the gauge field, $SU(2)_{Left}  $.
\f
p_a \phi (p)  \rightarrow {\cal D}_a \phi (p)  ) =  {\partial_a + A_a} \phi (p)  
\ff

\f
Z= < p_a^{i \ldots } |  \sum_\Gamma     ( \int dp_c^{p^{max}}  (  \int p^{max}  dp_d   )     d ( dp )  d (dr) d(ds)  \delta  (p+ w - q)   |   q_b^{j \ldots } >   
\ff     
Crucial to our theory is the incorporation of the gauge fields to turn global symmetries into local symmetries.
\f
p_a  \rightarrow D_a = p_a + \hat{A}_a
\ff
\item{} The next step is to introduce the right handed lepton fields.

\item{} Next we follow the same procedure to introduce the $SU(3)$ colours of quarks, along with an  $SU(3)$  gauge field.

\item{}Next, we introduce the $U(1)$ gauge fields.



\item{Next add the flavor symmetries}

We multiply the fermion fields, $\Psi $ and Higgs fields, $H $ by an $SU_"LEFT" (2)$. by putting the in fundamental representations of $SU_{Left}(2)$
\f
\phi \rightarrow H^{\  B}.  
\ff
The Hamiltonian constraint just changes by  putting it in a fundamental of $SU_{left} (2) .$,

\f
{\cal C} =  p^2 \rightarrow ,    \ \ \ \     | p_a p_b  I ^2 = p_b  g^{ab}  p_b
\ff

\item{Add colour symmetry. }
\f
P_{1  a}^b = {p^2} (  h^{a}_{b} + \frac{p^a p_b}{p^2}  )
\ff



 \item {Complete the gauge coupling}

Each trivalent and higher vertex involving gage fields comes from completion of the covariant derivative.  


\item{Impose the uv cutoffs  in the internal legs.}

Note that the integral will not be exactly gauge invariant, due to the cutoff. 


\end{itemize}

\section{Comments}

After we have put in the various interaction terms, what do we have?   We may note that we have put into the action all the terms
in the action for the standard model, as we would write it in momentum space.  Let us suppose we write these $N_{SM}$ terms as,   ${\cal L}_\alpha$.
\f
Z^{SM}= Tr    < p_a^{i}  |   \sum_\Gamma  \Pi_{ \Gamma}   \int dp_c^{p^{max}} \delta_{interactions} (\sum_{K a} ^ F ( p_{a L}   ) )  e^{\imath  \sum_{\alpha=1}^{N_{SM}}
{\cal L}_\alpha }
|p_b^{j } >  
\nonumber 
\ff
We can  make a diagrammatic expansion of by expanding around the bivalent terms, as usual.  We may first note that the
resulting series are not exactly the Feynman diagrams, because they do not have all the gauge invariance of Feynman diagrams.  For example, the cubic
vertex, gotten from grafting a gauge field only a quadratic propagator may have an independent coupling constant ; there is, it seems, no principle that ties the value
of the cubic graph to that of the quadratic graph.  

{\bf Dominance of gauge theories by gauge invariant terms at low energies.}

We are reminded of a informal idea in lattice gauge theory, sometimes called "multi-critical dynamics or "random dynamics",
which assumes that nature is governed by a theory with a finite, but large, uv cutoff, where it is described by a combination of
lattice gauge theory dynamics,  which are a random mixture of gauge and non-gauge invariant terms.  Then authors then claim that
the non-gauge  invariant couplings go away, when compared with the gauge invariant coupling such that in the infrared limit
the theory is governed by the gauge invariant sub-theory, alone\cite{Shenker, Nielsen}.

\section{Conclusions}

We close with a few remarks on on--going and future work.

\begin{enumerate}

\item{} We have now explored tentatively several corners of the framework of energetic causal  sets.  LLet us put it in some oersoective.

Counting the cosmological constant, $\Lambda = \frac{c^2}{R^2}$ as a fundamental observable, we have four fundamental, dimensional observables:      $\hbar, G, R, c $.  This is, famously, one too many.   But actually, there is one more: $N$,  the number of degrees of freedom.

Within any  proposed framework for quantum gravity, are then at least ten ways of reducing the description of our world to a simpler world, that still must be consistent, because it follows from a consistent set by taking one or more of the parameters to zero or infinity.

$ECS$ are one such proposed framework,  how are we doing?
First of alll, we have the well known corners:

\begin{itemize}

\item{Newtonian physics}

\item{Newtonian physics}
\f
c \rightarrow \infty,   \ \ \ \ \   \hbar \rightarrow 0, \ \ \ \ \ \ R \rightarrow \infty
\ff

\item{}Classical relativistic physics
\f
  \ \ \ \ \   \hbar \rightarrow 0, \ \ \ \ \ \ R \rightarrow \infty
\ff
\item{} Classical General Relativity
\f
\ \ \ \   \hbar \rightarrow 0, \ \
\ff

As the subject of Energetic Causal has developed we have scanned a ranges of theories.  In our first papers we studied
models of spacial relativistic particle based on a mechanism for the emergence of flat spacetime\cite{ECS1}-\cite{ECS4}.
Later we studied models of hidden variables in which we studied how non-relativistic quantum mechanics
emerges in a   limit $N \rightarrow \infty$
\cite{QMMV, VVQM}.  


These limits gave us non-relativistic quantum many-body theory.

In this paper we have broken through to the regieme  of $QFT$, although it appears we only recover lorentz invariance at low energies, compared to a 
fixed $uv$ cutoff.  

Clearly we have still some way way to go, most importantly we have to get off the  $G=0$ axis, to get gravity into the game.

One way to do this is to consider the $RL$ regieme in which $G$ and $\hbar $ both go to zero, with their ratio fixed.
\f
G \rightarrow 0, \ \ \ \ \hbar \rightarrow 0  \ \ \ \ \  \frac{\hbar}{G}  =   M_{Planck}^2. = const. 
\ff
\item{}Here we studied the case in which momentum space is flat, but the formalism we've developed may be easily extended to the
case. of non-linear momentum spaces. 
One way to do this is to deform the metric of momentum space in the hamiltonian constraints (\ref{C=0} )this forces new interactions amongst the
particles, as discussed in \cite{RL1}.  Whether this can be connected to a dynamical curvature on spacetime is
presently unknown.

\end{itemize}

\end{enumerate}

\section*{ACKNOWLEDGEMENTS}

II would like to thank Stefano Liberachi,   Laurent Freidel,  Clelia Verde, Ted Jacobson, Marina Cortes,  and Joao Magueijo for important conversations.  I am grateful to
Kai Smolin for the figures.
 This research was supported in part by Perimeter Institute for Theoretical Physics. Re- search at Perimeter Institute is supported by the Government of Canada through Industry Canada and by the Province of Ontario through the Ministry of Research and Innovation. This research was also partly supported by grants from NSERC, FQXi and the John Templeton Foundation.

\end{document}